# Raman Spectroscopy of magneto-phonon resonances in Graphene and Graphite

Sarah Goler

*NEST CNR-Istituto Nanoscienze and Scuola Normale Superiore Pisa, Italy*

Jun Yan

*Department of Physics, University of Maryland, College Park, Maryland 20742, USA*

Vittorio Pellegrini

*NEST CNR-Istituto Nanoscienze and Scuola Normale Superiore Pisa, Italy*

Aron Pinczuk

*Department of Physics and Department of Applied Physics and Applied Mathematics, Columbia University New York, USA*

**Abstract**

The magneto-phonon resonance or MPR occurs in semiconductor materials when the energy spacing between Landau levels is continuously tuned to cross the energy of an optical phonon mode. MPRs have been largely explored in bulk semiconductors, in two-dimensional systems and in quantum dots. Recently there has been significant interest in the MPR interactions of the Dirac fermion magneto-excitons in graphene, and a rich splitting and anti-crossing phenomena of the even parity $E_{2g}$ long wavelength optical phonon mode have been theoretically proposed and experimentally observed. The MPR has been found to crucially depend on disorder in the graphene layer. This is a feature that creates new venues for the study of interplays between disorder and interactions in the atomic layers. We review here the fundamentals of MRP in graphene and the experimental Raman scattering works that have led to the observation of these phenomena in graphene and graphite.

# 1. Introduction

Since the first measurement of the Raman spectrum of graphite [1], Raman scattering has become a popular characterization technique in carbon science and technology, widely used to probe disordered and amorphous carbons, fullerenes, nanotubes, diamonds, carbon chains, and poly-conjugated molecules [2]. After the first isolation of monolayer graphene [3], significant efforts have been made to investigate phonons, electron-phonon, and electron-electron interactions in graphene using Raman spectroscopy, and to understand how the number and orientation of the layers, the quality and types of edges, functional groups, as well as strain, doping and disorder affect the Raman processes [4, 5, 6, 7, 8, 9, 10, 11, 12, 13, 14, 15, 16, 17].

An interesting scenario can be anticipated under the influence of a magnetic field. In graphene the electronic transition energy between Landau levels (LLs) near the Dirac points sensitively depends on the magnetic field; it may occur for a particular field strength that the energy of the inter-LL excitations from the valence band to the conduction band matches the energy of the long-wavelength $E_{2g}$ optical phonon that in graphene and graphite occurs at around 1580 cm$_{-1}$ (the so-called G band). The large electron density of states in the LLs then dramatically enhances the electron-phonon interaction effects as compared with the case where the magnetic field is absent [5, 6]. This corresponds to a resonance between the optical phonon and the inter-LL electronic excitation, a condition called magneto-phonon resonance (MPR), predicted in Ref. [18].

MPR effects were explored before in bulk semiconductors [19, 20, 21], in two-dimensional semiconductor structures [22, 23, 24], and in quantum dots [25]. In graphene, the clear resolution of the coherent MPR effect requires weak broadening of the electronic transitions as compared with the electron-phonon coupling strength. This imposes severe restrictions on the sample quality. In fact, the observation of MPR had very different degrees of success in different samples (exfoliated graphene, epitaxial graphene, graphene flakes on graphite, bulk graphite). The corresponding results are reviewed in different subsections of Sec. 2.

The studies reviewed here demonstrate the rich physics that can be accessed by Raman probes in graphene and graphite samples subject to magnetic fields. Moreover, they allow for very precise characterizations of the material, such that many parameters related to electrons, phonons, and their interactions, can be determined with high accuracy.

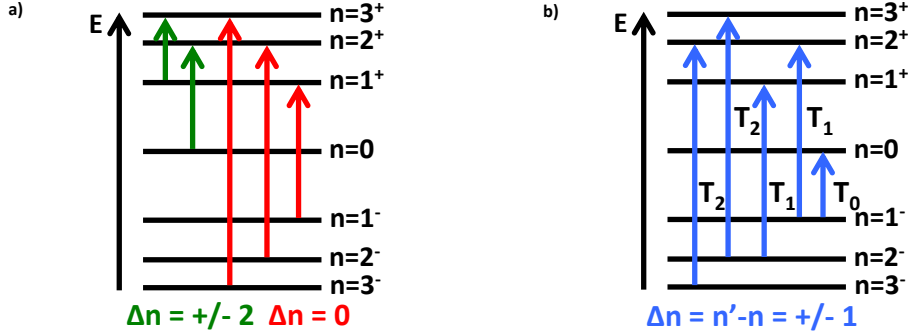

Figure 1: (color online) a) Inter-Landau level (LL) transitions, contributing to the Raman spectrum in the co-circular (red) and cross-circular (green) polarization configurations, shown by vertical arrows. The Landau levels in monolayer graphene are shown schematically by horizontal lines. b) Inter-LL transitions that are active in far-infrared spectroscopy experiments. Δn is the change in LL index associated to the transition. The axis labeled E corresponds to energy.

## 2. Magneto-phonon resonance

*2.1. Background*

Electron-phonon interaction produces corrections to lattice vibration frequencies. In graphene and graphite, such corrections have been shown to manifest themselves as the Kohn anomaly in the phonon dispersion [26], probed by the inelastic x-ray scattering [27, 28, 29], and as the dependence of the phonon frequency and decay rate on the charge carrier density [30], probed by Raman scattering as a function of external gate voltages [6, 5, 7, 8, 9]. Not surprisingly, these corrections are sensitive to the nature of the involved electronic excitations. The latter undergo a dramatic change when a strong magnetic field is applied. In graphene, a purely two-dimensional material, the electronic continuum collapses into a series of discrete Landau levels (which can be weakly broadened due to various scattering mechanisms). This results in a series of discrete couplings between optical phonons and electronic excitations, strongly enhanced when the two are in resonance, as predicted in Refs. [31, 32, 33]. Moreover, the discrete peaks corresponding to inter-LL excitations can be seen in Raman spectrum and identified by their dependence on the magnetic field [34], as anticipated in Refs. [35, 36].

For monolayer graphene in the Dirac approximation, the Landau levels can be labeled by a non-negative integer $n = 0, 1, \ldots$ and a sign "±", with the $n = 0$ level having no sign. Their energies are:

$$\epsilon_{n\pm} = \pm\sqrt{2n}\frac{\hbar v_F}{l_B}, \qquad (1)$$

where $v_F \approx 10^8$ cm/s is the electron velocity at the Dirac point, and $l_B = \left(\frac{e|B|}{\hbar c}\right)^{-\frac{1}{2}}$ is the magnetic length. For bilayer graphene, the levels are labeled as $0, 1, 2_\pm, 3_\pm, \ldots$ (we

use the convention of Ref. [37]). If one adopts the nearest-neighbor tight-binding model, parametrized by $v_F$ and the interlayer nearest-neighbor coupling matrix element $\gamma_1$, then in the two-band approximation the energies in bilayer graphene are given by:

$$\epsilon_{n\pm} = \pm \sqrt{n(n-1)} \frac{2\left(\frac{\hbar v_F}{l_B}\right)^2}{\gamma_1}, \quad n \geq 2, \tag{2}$$

and $\epsilon_0 = \epsilon_1 = 0$. The main selection rules for Raman scattering on inter-LL excitations can be straightforwardly deduced by noting that (i) a circularly polarized photon carries an angular momentum $m_z = \pm 1$, and (ii) an excitation between Landau levels with indices $n, n'$ carries angular momentum $m_z = \Delta n \equiv n' - n$. Thus, the transitions contributing to the Raman spectrum are either those with $\Delta n = 0$, seen in the co-circular polarization configuration (the incident and the scattered photons have the same polarization), or those with $\Delta n = \pm 2$, seen in the cross-circular configuration (the two photons have opposite polarizations), shown in Fig. 1(a). The actual calculation shows that the $\Delta n = 0$ transitions are more intense than those with $\Delta n = \pm 2$ [35, 36]. Similarly, the LL transitions involved in photon absorption are those between LLs with $\Delta n = \pm 1$ (Fig. 1(b)).

The properties of the doubly degenerate $E_{2g}$ optical phonon mode [see Fig. 2(a)] are probed by the Raman G peak at a frequency near 1580 cm$^{-1}$. The position and the shape of the peak are determined by the phonon spectral function:

$$S_{\pm}(\omega) = -\frac{1}{\pi} \text{Im} \frac{2\omega_{ph}}{\omega^2 - \omega_{ph}^2 + 2i\omega\gamma_{ph} - 2\lambda\omega_{ph}\Pi_{\pm}(\omega)}, \tag{3}$$

where $\omega_{ph}$ is the phonon frequency at zero magnetic field and zero doping, $2\gamma_{ph}$ is the phonon damping rate due to mechanisms other than electron-phonon coupling (e.g., anharmonicity), $\lambda$ is the dimensionless electron-phonon coupling constant defined in Ref. [30], and $\Pi(\omega)$ is the electronic polarization operator. The subscript "$\pm$" refers to the two circular polarization configurations in which the G peak is observed. Namely, the circular polarizations of the incident and scattered photons must be opposite to each other, then the emitted phonon is also circularly polarized, opposite to the incident photon; if the two photon polarizations are the same, the G peak is not observed (see Ref. [38] for a more detailed discussion of the selection rules).

When $\Pi(\omega)$ is a smooth enough function of the frequency, $S(\omega)$ is well approximated by a Lorentzian whose central frequency and half width at half maximum are given by the real and imaginary parts of the position of the pole $\tilde{\omega}$ of Eq. (3) in the complex plane of $\omega$. When $|\tilde{\omega} - \omega_{ph}| \ll \omega_{ph}$, $\tilde{\omega}$ can be found from the equation

$$\tilde{\omega} \approx \omega_{ph} - i\gamma_{ph} + \lambda\Pi(\tilde{\omega}). \tag{4}$$

Due to the smallness of $\lambda$ (of order $10^{-3}$), it is often sufficient to use the simplest approximation setting $\Pi(\tilde{\omega}) \to \Pi(\omega_{ph})$ in Eq. (4). This corresponds to the second-order perturbation theory in the electron-phonon coupling and is often sufficient to describe the experiments in zero magnetic field [6, 5, 7, 8, 9].

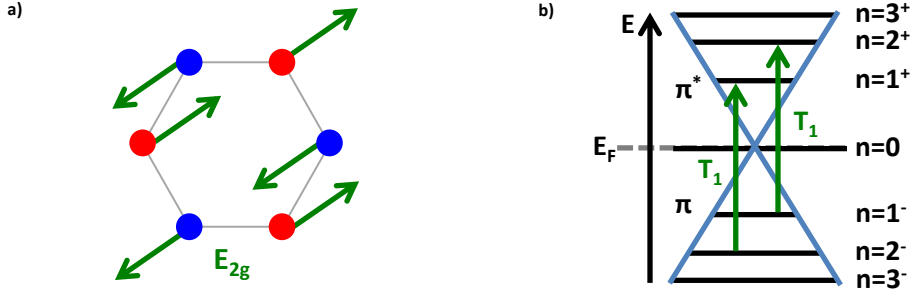

Figure 2: (color online) (a) The displacement pattern corresponding to the E$_{2g}$ optical phonon. (b) Inter-LL transitions which couple to the E$_{2g}$ phonon, shown by vertical arrows. The Landau levels in monolayer graphene are shown schematically by horizontal lines. The vertical axis labeled E is energy.

If a magnetic field $B$ is applied perpendicular to the graphene layer, $\Pi(\omega)$ is no longer a smooth function, and is given by the sum over transitions between Landau levels. In contrast to the electronic transitions contributing to Raman spectrum with index change $\Delta n = 0, \pm 2$, the transitions which couple to the E$_{2g}$ optical phonon are those with $\Delta n = \pm 1$, shown in Fig. 2(b). For monolayer graphene in the Dirac approximation, the polarization operator for both circular polarizations is given by [31, 32],

$$\Pi(\omega) = \sum_{n=0}^{\infty} \frac{\left(\frac{\omega^2 \Omega_0^2}{\Omega_n}\right)}{\omega^2 + 2i\omega\gamma_{el} - \Omega_n^2}, \quad (5)$$

where $\hbar\Omega_n = \epsilon_{(n+1)^+} - \epsilon_{n^-} = \epsilon_{n^+} - \epsilon_{(n+1)^-}$ is the transition energy, and $2\hbar\gamma_{el}$ is the transition broadening due to various electronic scattering processes. For bilayer graphene or bulk graphite, the four-band model has to be employed for the calculation of $\Pi(\omega)$ in the frequency region of interest [33, 38], so no simple analytical expression is available. Still, the selection rule is the same, and the structure of $\Pi(\omega)$ is analogous. For bulk graphite, an additional integration over a continuous quantum number $k_z$ (the component of the wave vector perpendicular to the layers) is required, so the discrete resonances are smeared into singularities in the joint density of states [38].

As the magnetic field is varied, various transitions $\Omega_n$ may match the phonon frequency $\omega_{ph}$, which corresponds to the resonance between the optical phonon and the inter-LL electronic excitation, the so-called magneto-phonon resonance (MPR). Near the resonance, $\Pi(\omega)$ is no longer smooth, and second-order perturbation theory is insufficient to describe the interaction behaviors. Still, if $\gamma_{el}$ is not too small (note that typically, $\gamma_{el} \gg \gamma_{ph}$), or the detuning |$\Omega_n$- $\omega_{ph}$| is large enough, $S(\omega)$ may be approximated by a single Lorentzian, but the pole position is obtained only after several iterations of Eq. (4), where each subsequent approximation is obtained by using the previous one in the argument of $\Pi(\omega)$. Convergence after a few iterations corresponds to some high, but finite, order of the perturbation theory.

However, if both $\gamma_{el}$ and $\gamma_{ph}$ are small enough, at resonance the spectral function is no longer a single Lorentzian peak, but consists of two peaks. The two peaks correspond to coherent superpositions of the two coupled excitations – the optical phonon and the discrete electronic excitation. The two frequencies, corresponding to two split poles of Eq. (3) are approximately given by the solutions of the corresponding quadratic equation:

$$\tilde{\omega}_{1,2} = \frac{\omega_{ph} + \Omega_n}{2} - i\frac{\gamma_{ph} + \gamma_{el}}{2} \pm \frac{1}{2}\sqrt{2\lambda\Omega_0^2 + [\Omega_n - \omega_{ph} - i(\gamma_{el} + \gamma_{ph})]^2}. \tag{6}$$

The presence of two peaks cannot be captured in any finite order of the perturbation theory and corresponds to the regime of strong coupling between the optical phonon and the inter-LL excitation. Such a regime occurs only when the coherent coupling between the two excitations overcomes the damping, which tends to destroy their coherent superposition: from Eq. (6), exactly at resonance, $\omega_{ph} - \Omega_n$, the square root becomes purely imaginary (so the splitting vanishes). This does not happen when

$$\gamma_{el} - \gamma_{ph} < \frac{\sqrt{2\lambda}\omega_{ph}}{\sqrt{n} + \sqrt{n+1}}, \tag{7}$$

which can be considered a criterion for the strong-coupling regime.

*2.2. Magneto-phonon resonance in exfoliated monolayer graphene*

Several attempts to trace the MPR effects in graphene flakes on Si/SiO$_2$ substrates were made. The results of Ref. [39] suggest that some changes in the lineshape of the *E$_{2g}$* mode occur upon varying the magnetic field across 5 T (the value of *B* at which the resonance *T$_1$*, $\Omega_1 = \omega_{ph}$ occurs), but no clear resolution of the MPR was achieved (see Fig. 3a)). Ref. [40] has reported a splitting and not a simple shift of the *E$_{2g}$* phonon in the Raman spectrum at B = 11 T, likely obtained on quasineutral graphene. However, this value of *B* does not correspond to any resonance that could be expected with a reasonable value of $v_F$.

We believe the main reason for the failure to observe the MPR at low-magnetic fields in exfoliated monolayer graphene is due to the large broadening of the transitions which smears the effect. This conjecture is supported by far-infrared absorption spectra obtained by Jiang et al. [41] (see Fig. 3b)). The spectra at low magnetic field do not resolve the *T$_1$* transitions probably due to its large linewidth. The large broadening of the *T$_1$*, however, does not seem to be a fundamental obstacle, since extremely narrow transitions have been observed in graphene flakes on graphite, discussed in the following subsection. We do not know the exact mechanism responsible for the smearing of the MPR effect in exfoliated graphene,[1] but we do not see any fundamental reason why such narrow transitions cannot be observed there as well. The primary challenge seems to be mostly technological: to prepare high-quality samples and to isolate them from the environment. So, the quest for the observation of the magneto-phonon resonance in exfoliated graphene is still open.

We should say that recent Raman experiments on exfoliated graphene performed in high-magnetic field facilities have indeed observed several MPR phenomena at magnetic fields above 10-15 Tesla [42].

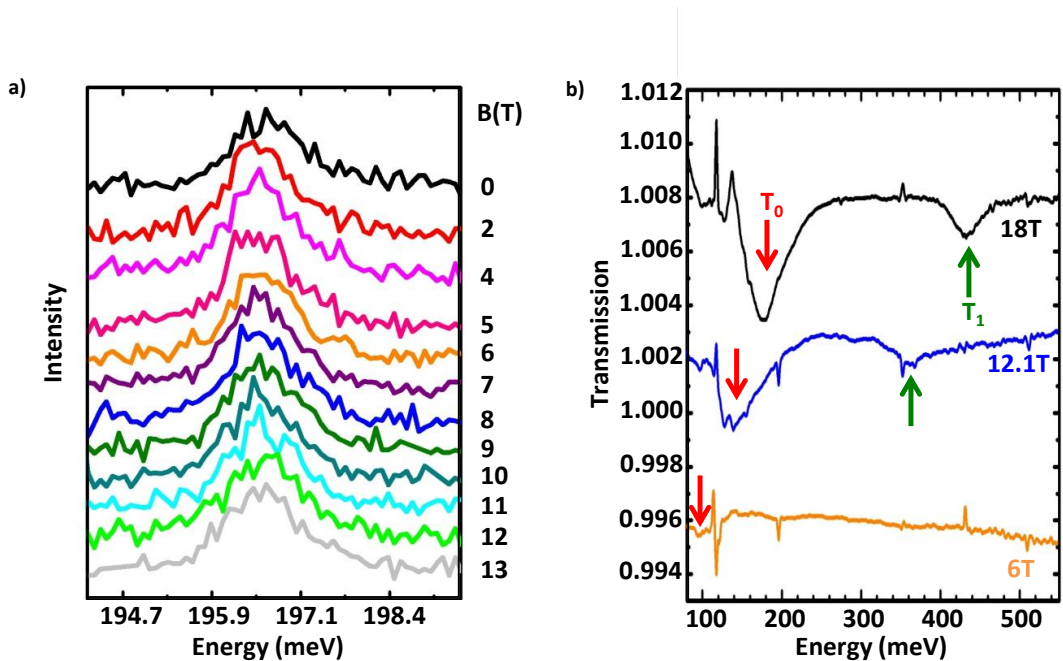

Figure 3: (color online) a) Magnetic field dependence of the *G* phonon band in exfoliated monolayer graphene on a Si/SiO2 substrate obtained by Raman scattering at a temperature of ≈ 2K . b) Representative far-infrared absorption spectra at 4.2K. (from [41])

*2.3. Magneto-phonon resonances of graphene in graphite*

Shortly after the early unsuccessful attempts to observe the MPR in exfoliated graphene, it was discovered that the best graphene is in fact hidden in graphite,

---

[1] Spatial fluctuations of the charge density, pervasive in exfoliated graphene layers de- posited on SiO2 substrates, do not seem to be the main reason for the resolution failure. Indeed, the usual inhomogeneous broadening of the phonon frequency by the density fluctuations [6, 5] is quenched in a magnetic field due to the quantization of the electronic spectrum [48]. As for the electronic inter-LL transitions, the fluctuating charge density mainly results in a spatially local shift of the electronic bands as a whole, which does not affect the transition energies. To inhomogeneously broaden the inter-LL transitions, local fluctuations of $v_F$ are required.

which is probably less influenced by various environmental perturbations [39].

The resonant nature of the electron-phonon interaction in graphene in high magnetic fields was clearly demonstrated in multilayer epitaxial graphene grown on the carbon face of a SiC substrate [43]. The electronic bands of this material are similar to those of a single layer, because the stacking of layers is turbostratic and not Bernal-type, which is typical of the usual bulk graphite and multilayer graphene [44, 45, 46, 47]. Upon the application of magnetic fields up to 33 T at a temperature of 4.2 K, the Raman $G$ band was found to consist of two components; one was insensitive to the magnetic field, and the other one exhibited pronounced variations of both the position and width, shown in Fig. 4. Avoided crossings were seen at values of $B$ corresponding to the resonances $\Omega_n = \omega_{ph}$. Variations of both the position and width of the peak could be reproduced by taking Re$\widetilde{\omega}$, Im $\omega$ from the iterative solution of Eq. (4). The values of $\omega_{ph}$ = 1586.5 cm$^{-1}$, $v_F$ = 1.02 × 10$^8$ cm/s, $\lambda$ = 4.5 × 10$^{-3}$. The extracted value of $\gamma_{el}$ = 90 cm$^{-1}$ satisfies the strong-coupling criterion, Eq. (7), only for the strongest resonance $\Omega_0 = \omega_{ph}$. This resonance occurs at 28 T, only the low-$B$ side of this resonance was observed in the experiment due to the limited range of magnetic fields.

The MPR effect was found to be much more pronounced in inclusions of monolayer graphene in bulk graphite [48] where the occurrence of the coherent coupling between the $T_1$ and $E_{2g}$ occurring near 5T was achieved. We stress that the existence of high-quality decoupled graphene flakes in bulk graphite (presumably, on the surface) was suggested by several theoretical and experimental studies [44, 49, 50]. By scanning a Kish graphite sample with a moderate spatial resolution of approximately 80 μm at a temperature of about 2 K, one could identify regions which exhibited graphene-like MPR behavior, in contrast to other regions which did not. The Raman $E_{2g}$ band from the graphene-like regions in a magnetic field consisted of two components, one weakly sensitive to the field, and the other one exhibiting a pronounced MPR effect with a clear anti-crossing behavior. Focusing on the resonance $\Omega_1 = \omega_{ph}$ at B = 5 T (see Fig. 5), and fitting the peak position and width with Eq. (6), the authors [48] could extract the parameters as $\omega_{ph}$ = 1584cm$^{-1}$, $v_F$ = 1.03×10$_8$ cm/s, $\lambda$ = 6.36×10$^{-3}$. The value of $\gamma_{el}$ = 26 cm$^{-1}$ was about 3 times lower than that observed on other samples [43, 51, 49]. Thanks to this small value, the strong-coupling criterion (7) was well satisfied for the $\Omega_1 = \omega_{ph}$ resonance. The fact that the split peaks are not seen in the narrow region 4.9 T < $B$ < 5.1 T does not mean that they are absent: at resonance each peak gets a 50% share of $\gamma_{el}$, so because of the broadening the height is too small and is beyond the sensitivity of the spectrometer.

Further support of the interpretation linking the Dirac-like MPR behavior to decoupled monolayer graphene flakes in graphite comes from more recent Raman studies on natural graphite using an optical set-up with monomode and multimode fibers able to reach a spatial resolution of 1μm [34, 52]. These experiments, performed at 4.2 K and up to a magnetic field of 28 T, have also clearly identified the MPR anti-crossing of the $E_{2g}$ phonon with inter- LL transitions $n^- \rightarrow (n+1)^+$, $(n+1)^- \rightarrow n^+$. Moreover, the anticrossing with these two series, observed in the opposite cross-circular polarization configurations, occurred at different values of the magnetic field. This is a manifestation of the electron-hole asymmetry, which results in $\epsilon_{(n+1)+} - \epsilon_{n-} \neq \epsilon_{n+} - \epsilon_{(n+1)-}$.

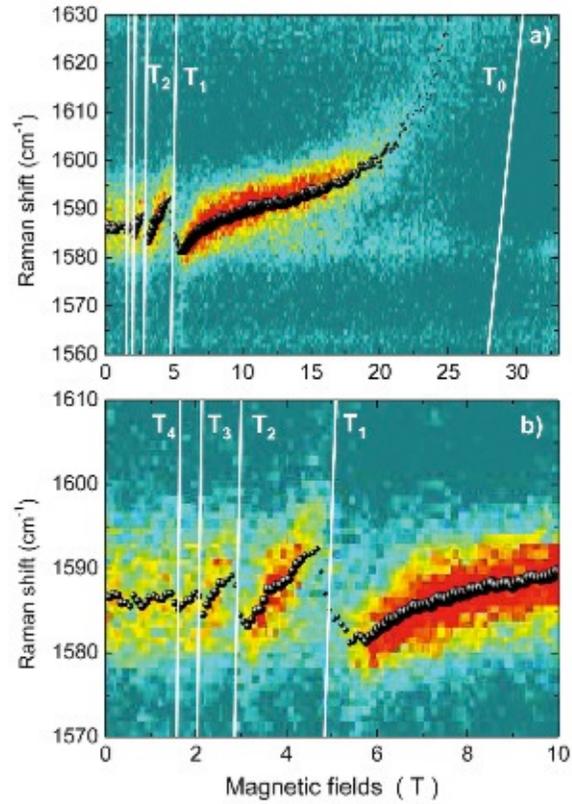

Figure 4: (color online) (a) Color map of the magneto-oscillatory component of the Raman *G* band as an acfunction of the magnetic field. The extracted peak position of this line is shown with full dots, whose size is proportional to the line amplitude. White solid lines $T_n$ represent the frequencies $\Omega_n$ of inter-LL transitions. (b) Zoom of (a) on the B < 10 T range. (Reprinted from Ref. [43].)

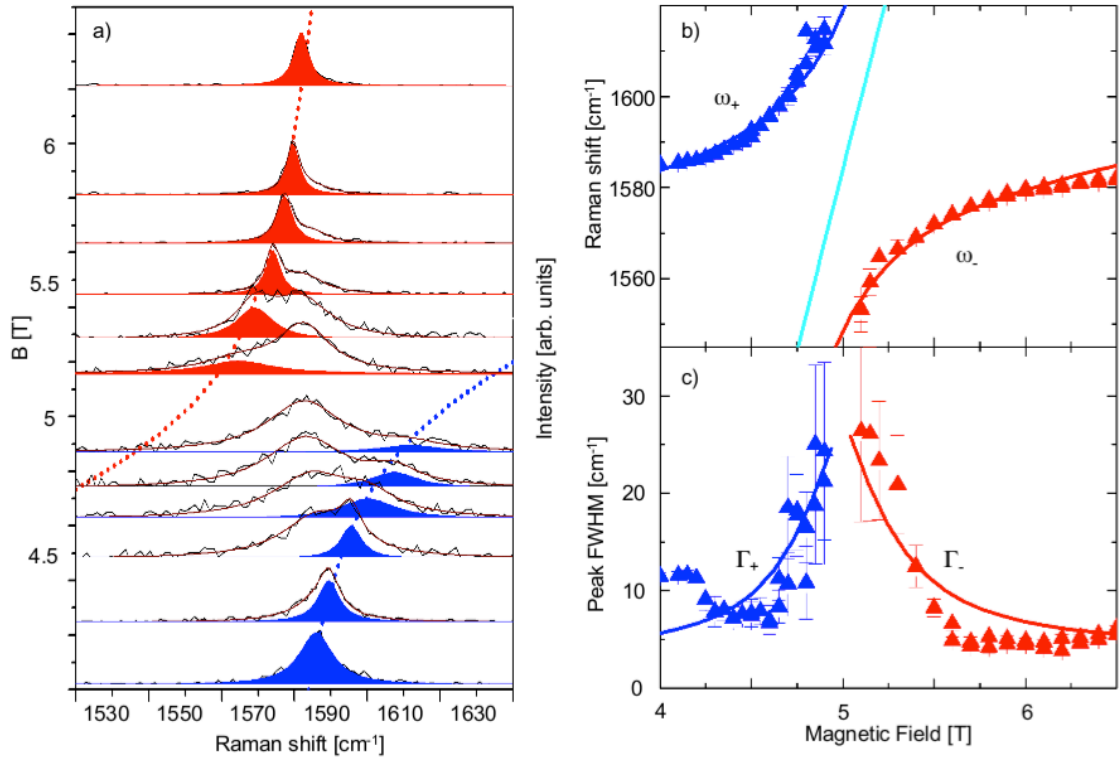

Figure 5: (color online) (a) Raman spectra for magnetic fields in the range 4.2 T < B < 6.4 T. The spectra are shifted vertically so that each red and blue component taken at a given B has its peak aligned at this B. The spectral component that corresponds to the two coupled modes is highlighted. The dotted lines represent the fit using Eq. (6). (b) Mode energy of the coupled modes versus B. The blue and red triangles represent the experimental data. The fit using Eq. (6) is shown by blue and red lines. The cyan line represents $\Omega_1$. (c) The same for the full width of half maximum of the spectral peaks. (Reprinted from Ref. [48].)

Finally, the results presented above stimulated further interests on bulk graphite at large magnetic fields. In Ref. [53] the Raman $E_{2g}$ peak of natural graphite was observed to shift and split as a function of magnetic field up to 45T, due to the magnetically tuned coupling with the K- and H-point inter- LL excitations. The authors in Ref.[54] used magneto-Raman scattering to study purely electronic excitations and the electron-phonon coupling in bulk graphite up to 28 T. They observed the magneto-phonon effect involving the $E_{2g}$ optical phonon and K-point inter-LL excitations with $\delta n = \pm 1$ and successfully interpreted the results within a model taking into account the full $k_z$ dispersion.

## 3. Conclusions

In this paper, we have reviewed the main theoretical and experimental results obtained so far concerning the influence of an external magnetic field on Raman spectra of graphene and graphite. In spite of significant progress, many issues remain unsettled. Thus, the field is still open for further theoretical and experimental works. MPRs in high-quality exfoliated graphene at low magnetic fields and in graphene and bilayer graphene on SiC are among the topics to be addressed in future investigations.

## 4. Acknowledgements

We wish to acknowledge Denis Basko for the precious help in writing the manuscript. We also thank Andrea Ferrari for several useful discussions and Erik A. Henriksen for sending us the far-infrared absorption data in graphene. Work in Pisa was supported by the Italian Ministry of Education, University, and Research (MIUR) through the program "FIRB - Futuro in Ricerca 2010" Grant No. RBFR10M5BT ("PLASMOGRAPH: plasmons and terahertz devices in graphene"). The work by A.P. is supported by ONR (N000140610138 and Graphene Muri), NSF (CHE-0641523), and NYSTAR.